\documentclass[a4paper]{spie}  

\newcommand{\cii}{[C\,{\sc ii}]\,}
\newcommand{\oiii}{[O\,{\sc iii}]\,}
\newcommand{\nii}{[N\,{\sc ii}]\,}
\newcommand{\ciii}{[C\,{\sc iii}]\,1907\AA + C\,{\sc iii}]\,1909\AA\,}

\usepackage{amsmath,amsfonts,amssymb}
\usepackage{graphicx}
\usepackage[colorlinks=true, allcolors=blue]{hyperref}

\title{Large format imaging spectrograph for the Large Submillimeter Telescope (LST)}

\author[a,b]{Kotaro Kohno}
\author[c,d]{Ryohei Kawabe}
\author[e]{Yoichi Tamura}
\author[f,g]{Akira Endo}
\author[h,f]{Jochem J. A. Baselmans}
\author[h,f]{Kenichi Karatsu}
\author[i,j,k]{Akio K. Inoue}
\author[l]{Kana Moriwaki}
\author[l]{Natsuki H. Hayatsu}
\author[l,m,b]{Naoki Yoshida}
\author[a]{Yuki Yoshimura}
\author[a]{Bunyo Hatsukade}
\author[n,a]{Hideki Umehata}
\author[c]{Tai Oshima}
\author[o,a]{Tatsuya Takekoshi}
\author[e]{Akio Taniguchi}
\author[p]{Pamela D. Klaassen}
\author[q]{Tony Mroczkowski}
\author[r]{Claudia Cicone} 
\author[s,t]{Frank Bertoldi}
\author[u,v]{Helmut Dannerbauer}
\author[w]{Tomoka Tosaki}

\affil[a]{Institute of Astronomy, School of Science, The University of Tokyo, 2-21-1 Osawa, Mitaka, Tokyo 181-0015, Japan}
\affil[b]{Research Center for the Early Universe, School of Science, The University of Tokyo, 7-3-1 Hongo, Bunkyo, Tokyo 113-0033, Japan}
\affil[c]{National Astronomical Observatory of Japan, 2-21-1 Osawa, Mitaka, Tokyo 181-8588, Japan}
\affil[d]{The Graduate University for Advanced Studies (SOKENDAI), 2-21-1 Osawa, Mitaka, Tokyo 181-8588, Japan}
\affil[e]{Division of Particle and Astrophysical Science, Graduate School of Science, Nagoya University, Nagoya 464-8602, Japan}
\affil[f]{Faculty of Electrical Engineering, Mathematics and Computer Science, Delft University of Technology, Delft, The Netherlands}
\affil[g]{Kavli Institute of NanoScience, Faculty of Applied Sciences, Delft University of Technology, Delft, The Netherlands}
\affil[h]{SRON -- Netherlands Institute for Space Research, Utrecht, The Netherlands}
\affil[i]{Department of Physics, School of Advanced Science and Engineering, Faculty of Science and Engineering,
Waseda University, 3-4-1 Okubo, Shinjuku, Tokyo 169-8555, Japan}
\affil[j]{Waseda Research Institute for Science and Engineering, Faculty of Science and Engineering, Waseda University, 3-4-1 Okubo, Shinjuku, Tokyo 169-8555, Japan}
\affil[k]{Department of Environmental Science and Technology, Faculty of Design Technology, Osaka Sangyo University, 3-1-1 Nakagaito, Daito, Osaka 574-8530, Japan}
\affil[l]{Department of Physics, School of Science, The University of Tokyo, 7-3-1 Hongo, Bunkyo, Tokyo 113-0033, Japan}
\affil[m]{Kavli Institute for the Physics and Mathematics of the Universe (WPI), UT Institutes for Advanced Study, The University of Tokyo, 5-1-5 Kashiwanoha,
Kashiwa, Chiba 277-8583, Japan}
\affil[n]{RIKEN Cluster for Pioneering Research, 2-1 Hirosawa, Wako, Saitama 351-0198, Japan}
\affil[o]{Kitami Institute of Technology, 165 Koen-cho, Kitami, Hokkaido 090-8507, Japan}
\affil[p]{UK Astronomy Technology Centre, Royal Observatory Edinburgh, Blackford Hill, Edinburgh EH9 3HJ, UK}
\affil[q]{European Southern Observatory (ESO), Karl-Schwarzschild-Strasse 2, Garching 85748, Germany}
\affil[r]{Institute of Theoretical Astrophysics, University of Oslo, P.O. Box 1029, Blindern, 0315 Oslo, Norway}
\affil[s]{Max-Planck-Institut f\"{u}r Radioastronomie, Auf dem H\"{u}gel 69, D-53121 Bonn, Germany}
\affil[t]{Argelander-Institut f\"{u}r Astronomie, University at Bonn, Auf dem H\"{u}gel 71, D-53121 Bonn, Germany}
\affil[u]{Instituto de Astrof\'{i}sica de Canarias, E-38205 La Laguna, Tenerife, Spain}
\affil[v]{Universidad de La Laguna, Departamento de Astrof\'{i}sica, E-38206 La Laguna, Tenerife, Spain}
\affil[w]{Joetsu University of Education, Yamayashiki-machi, Joetsu, Niigata 943-8512, Japan}

\authorinfo{Further author information: (Send correspondence to K.K.)\\K.K.: E-mail: kkohno@ioa.s.u-tokyo.ac.jp}

\begin{document} 
\maketitle

\begin{abstract}
We present a conceptual study of a large format imaging spectrograph for the Large Submillimeter Telescope (LST) and the Atacama Large Aperture Submillimeter Telescope (AtLAST). Recent observations of high-redshift galaxies indicate the onset of earliest star formation just a few 100 million years after the Big Bang (i.e., $z$ = 12--15), and LST/AtLAST will provide a unique pathway to uncover spectroscopically-identified ``first forming galaxies’’ in the pre-reionization era, once it will be equipped with a large format imaging spectrograph. We propose a 3-band (200, 255, and 350 GHz), medium resolution ($R$ = 2,000) imaging spectrograph with $\sim$1.5 M detectors in total based on the KATANA concept (Karatsu et al.~2019), which exploits technologies of the integrated superconducting spectrometer (ISS) and a large-format imaging array. A 1-deg$^2$ drilling survey (3,500 hr) will capture a large number of \oiii 88 $\mu$m (and \cii 158 $\mu$m) emitters at $z$ = 8--9, and constrain \oiii luminosity functions at $z>12$.
\end{abstract}


\section{INTRODUCTION}
\label{sec:intro}  

\begin{figure} [ht]
   \begin{center}
   \begin{tabular}{c} 
   \includegraphics[height=9cm]{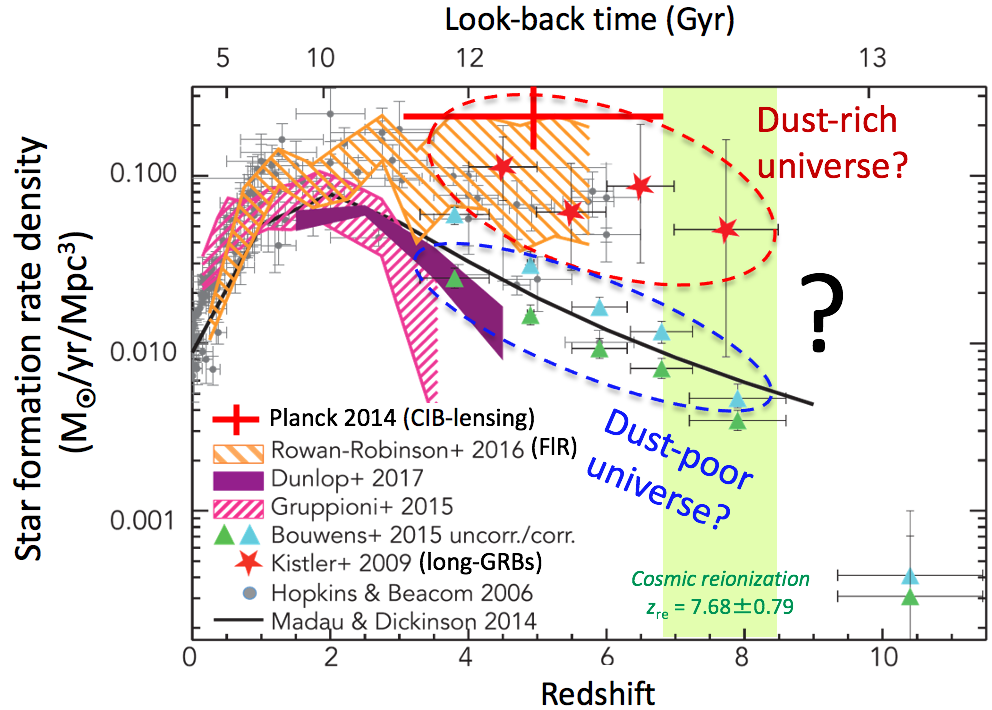}
   \end{tabular}
   \end{center}
   \caption[example] 
   { \label{fig:CSFRD} 
The redshift evolution of the cosmic star-formation rate density, illustrating two inconsistent trends of measurements at $z = 4 - 8$, i.e., ``dust-rich, actively star-forming universe'' (suggested by {\it Herschel} - FIR, long-$\gamma$-ray bursts, and cosmic infrared background-light (CIB) lensing analysis with {\it Planck}) and ``dust-poor, calm early universe'' (suggested by observations of Lyman break galaxies with dust-extinction corrections). }
   \end{figure}

Recent multi-wavelength surveys of distant galaxies have revealed that the majority of the star-forming activities at a redshift $z =$ 1--3, where the cosmic star formation rate densities peak, is obscured by dust, emphasizing importance of the study of dust-obscured activities in the universe. 
However, the roles of the dust-obscured star-formation beyond the redshift of $z>$ 4--8 remain unclear, because different measurement techniques result in different indications, as shown in Figure \ref{fig:CSFRD}. 
For instance, ALMA observations of Lyman break galaxies in {\it Hubble} Ultra Deep Field suggest that the dust obscured star formation plays rather minor roles among such star-forming galaxies selected by rest-frame ultraviolet (UV) radiation\cite{Bouwens2016,Bouwens2020}. On the other hand, {\it Herschel} observations of ``red'' SPIRE sources suggest elevated star-formation rate densities up to $z\sim6$ \cite{Rowan-Robinson2016}, which is also supported by recent ALMA observations \cite{Gruppioni2020}.
%
Latest ALMA studies also demonstrate the presence of sub/millimeter-selected galaxies without any significant counterpart seen in the optical and near-infrared: {\it HST}-dark galaxies \cite{Fujimoto2016,Wang2019,Yamaguchi2019,Umehata2020,Smail2020}.
Although such a class of galaxies has been known from the beginning of the discovery of the submillimeter galaxies (SMGs), recent ALMA observations unveil much fainter sub/millimeter sources ($S_{\rm 1mm}\sim$ a few 10--100 $\mu$Jy), which were unable to isolate before due to the source confusion limit of submillimeter-wave survey facilities like SCUBA/SCUBA2 on JCMT 15-m telescope and AzTEC on ASTE 10-m telescope. And more importantly, such faint sub/millimeter galaxies are much more ubiquitous\cite{Wang2019} and therefore responsible for the bulk of the cosmic infrared background-light (CIB)\cite{Fujimoto2016,Hatsukade2018,Jorge2020}. The importance of {\it HST}-dark but IRAC-detected (a.k.a. {\it H}-dropout) galaxies as a key tracer of the early phases of massive galaxy formation at $z\sim$ 4--6
has been discussed 
\cite{Wang2019,Zhou2020}, but the difficulty of obtaining spectroscopic redshifts for such {\it HST}-dark galaxies hampers the physical characterization of these sources. 
After pioneering spectral scan observations of the {\it Hubble} Deep Field North using IRAM Plateau de Bure Interferometer \cite{Decarli14}, ALMA and NOEMA have been used to search for millimeter-wave line emitters\cite{Walter16,Aravena19,Yamaguchi17,Gonzalez17,Tamura14,Gowardhan19,Kaasinen20} without any priors in the optical/near-infrared bands, but the total surveyed area has been (and will be) limited to tiny patches of the sky because of the narrowness of the ALMA/NOEMA field-of-view.
All of this progress motivates us to design a wide-area spectroscopic survey of dust-enshrouded galaxies in a more systematic way \cite{Geach2019}.

Recent discoveries of star-forming galaxies at $z$ = 8--11\cite{Oesch2016,Hashimoto2018,Tamura2019} and candidate passive galaxies at $z\sim6$\cite{Mawatari2020} indicate the onset of the earliest star formation just a few 100 million years after the Big Bang, i.e., $z \sim$ 12--15. 
This prompts us to establish a methodology to uncover a statistically large number of spectroscopically-identified, ``first forming galaxies’’ during the epoch of reionization (EoR) and beyond, i.e., the pre-reionization era/cosmic dawn. Although both ALMA and {\it JWST}/NIRSpec can detect \oiii 88 $\mu$m and \ciii lines, respectively, at $z>$ 10--15 with a reasonable observing time, they are not optimized for a wide-area survey to uncover such candidate sources. A $>$100 deg$^2$ near-infrared imaging survey at $\lambda$ = 2--5 $\mu$m with a modest depth ($\sim$27 $m_{\rm AB}$) would be the optimum option to find spectroscopic follow-up targets at $z\sim15$ (Inoue et al.) but the Roman Space Telescope will have no survey capability at $\lambda>2$ $\mu$m. 

Here we argue that next-generation large submillimeter single-dish telescopes optimized for wide-area, broad-band spectral coverage surveys, such as the Large Submillimeter Telescope (LST)\cite{Kawabe2016}\footnote{\url{https://www.lstobservatory.org}} and the Atacama Large Aperture Submillimeter Telescope (AtLAST)\cite{Klaassen2020}\footnote{\url{https://atlast-telescope.org}}, will provide a unique pathway for that purpose once these telescopes are equipped with a large format imaging spectrograph\cite{Kohno2019}. 
In this paper, we present a conceptual study of a 3-band imaging spectrograph, which specifically aims for the \oiii 88 $\mu$m and the \cii 158 $\mu$m tomography at $z$ = 4--8 (with \cii) and $z$ = 8--16 (with \oiii), based on the KATANA concept (Karatsu et al.~2019)\footnote{\url{https://agenda.infn.it/event/15448/contributions/95630/}} using the technologies of the integrated superconducting spectrometer (ISS)\cite{Endo2019NA,Karkare2020} and a large-format imaging array like A-MKID\cite{Baselmans2018}. Our goal here is not to propose a very detailed specification of the observing instrument, but to provide a conceivable observing instrument case based on a set of science requirements; it will give some insights and implications for further investigations of the observing instrument and technologies for LST/AtLAST. A brief overview and updates of the LST project are also given. 

\section{The Large submillimeter telescope}

The LST project was originally discussed as the planning of a next-generation telescope for sub/millimeter-wavelengths that could inherit both the large collecting area of the Nobeyama Radio Observatory (NRO) 45-m telescope\cite{Ukita1994} and the submillimeter capabilities of Atacama Submillimeter Telescope Experiment (ASTE) 10-m telescope \cite{Ezawa2008}. 
The current major specifications of the LST, along with the conceptual design and key science behind have been summarized in Kawabe et al. (2016)\cite{Kawabe2016}. In brief, the LST will be a 50-m diameter high-precision (45 $\mu$m rms) telescope for wide-area imaging and spectroscopic surveys with a field-of-view of $>$0.5 deg $\phi$ primarily focusing on the 70--420 GHz frequency range, with a capability for high frequencies up to 1 THz using an inner high-precision surface. A novel concept of a millimeter wavefront sensor that allows real-time sensing of the surface, which will be a key to establish ``millimetric adaptive optics (MAO)'', has been proposed by Tamura, Y., et al.\cite{Tamura2020mao}. Statistical approaches to efficient atmospheric noise removal for submillimeter-wave spectroscopy have been proposed and implemented by Taniguchi, A., et al. for e.g., NRO 45-m telescope\cite{Taniguchi2020} and DESHIMA on ASTE\cite{Taniguchi2020spie}.

\begin{figure} [ht]
   \begin{center}
   \begin{tabular}{c} 
   \includegraphics[height=10cm]{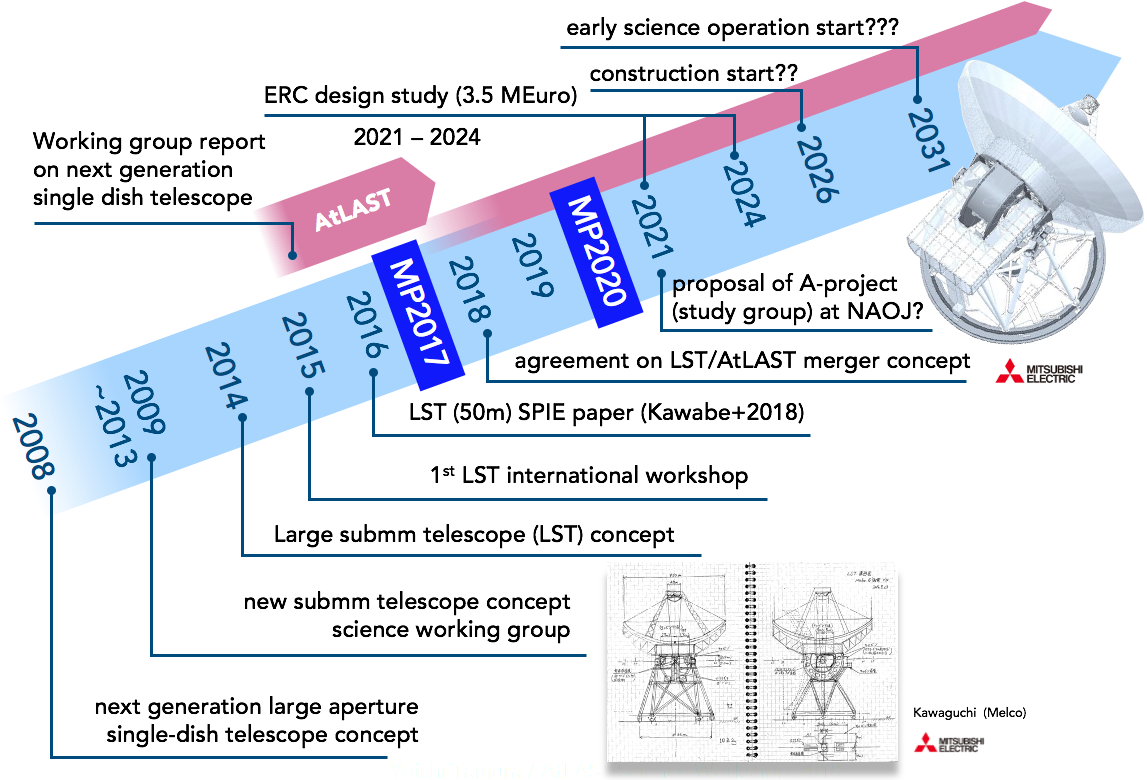}
   \end{tabular}
   \end{center}
   \caption[example] 
   { \label{fig:timeline} 
LST project timeline and milestones.}
   \end{figure}

One of the recent major milestones of the LST project is the master plan 2020 (MP2020) led by the Science Council of Japan (SCJ), which aims to set the list of high priority large academic research projects in Japan. The LST project was invited to give a presentation at two MP2020 symposia in September 2018 and January 2019, which were organized by the astronomy and astrophysics sub-committee of SCJ, to discuss the progress of the project since the previous master plan activity (MP2017) and the status of the international collaboration including the EU-led AtLAST project. A support letter from the AtLAST community was highly appreciated. Now the LST is formally listed as one of the large academic projects in the astronomy and astrophysics field in MP2020, which was announced in January 2020. During these activities, a merger between the AtLAST and LST projects has been intensively discussed. After the success of the 3.5 M Euro ERC program for the AtLAST design study in 2021--2024\cite{Klaassen2020}, we anticipate the merger in 2024 where the design study will end. There are some apparent inconsistencies of telescope specifications such as the target surface accuracy and the field of view, which are expected to be discussed further during the AtLAST design study in coming years. A next step forward in the LST would be a proposal to National Astronomical Observatory of Japan (NAOJ) to launch a study group of the LST under a support from the community in Japan. Figure \ref{fig:timeline} displays the LST timelines and milestones, along with those of the AtLAST. 

   \begin{figure} [htb]
   \begin{center}
   \begin{tabular}{c} 
   \includegraphics[height=7.3cm]{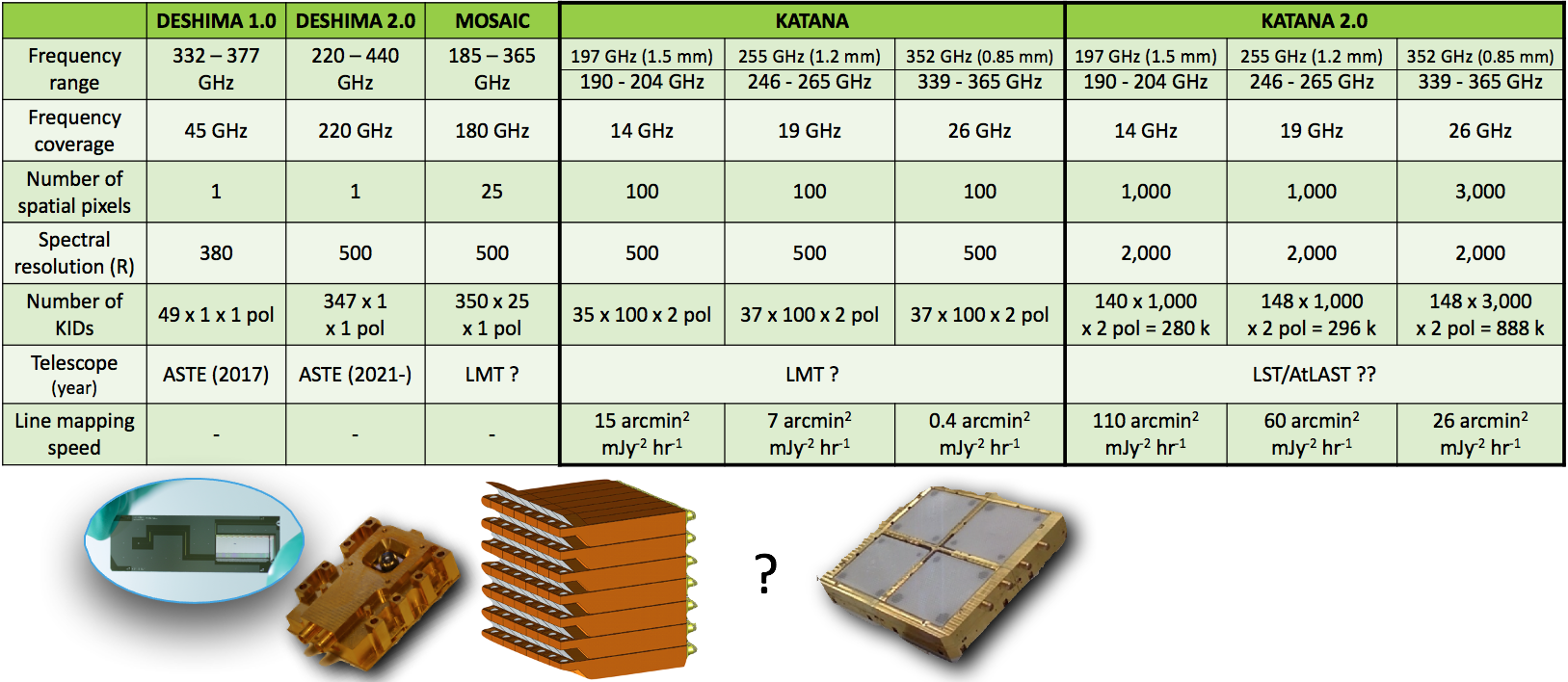}
   \end{tabular}
   \end{center}
   \caption[example] 
   { \label{fig:deshima2katana} 
Summary of the major specifications of DESHIMA, MOSAIC, and the proposed 3-band imaging spectrograph based on the KATANA concept, which will exploit technologies of the integrated superconducting spectrometer (DESHIMA) and a large-format imaging array (like A-MKID). Pictures of DESHIMA and A-MKID, along with the proposed array configuration of MOSAIC, are inserted.}
   \end{figure} 

\section{From DESHIMA to KATANA: 3-band imaging spectrograph} 

\begin{figure} [htb]
   \begin{center}
   \begin{tabular}{c} 
   \includegraphics[height=12cm]{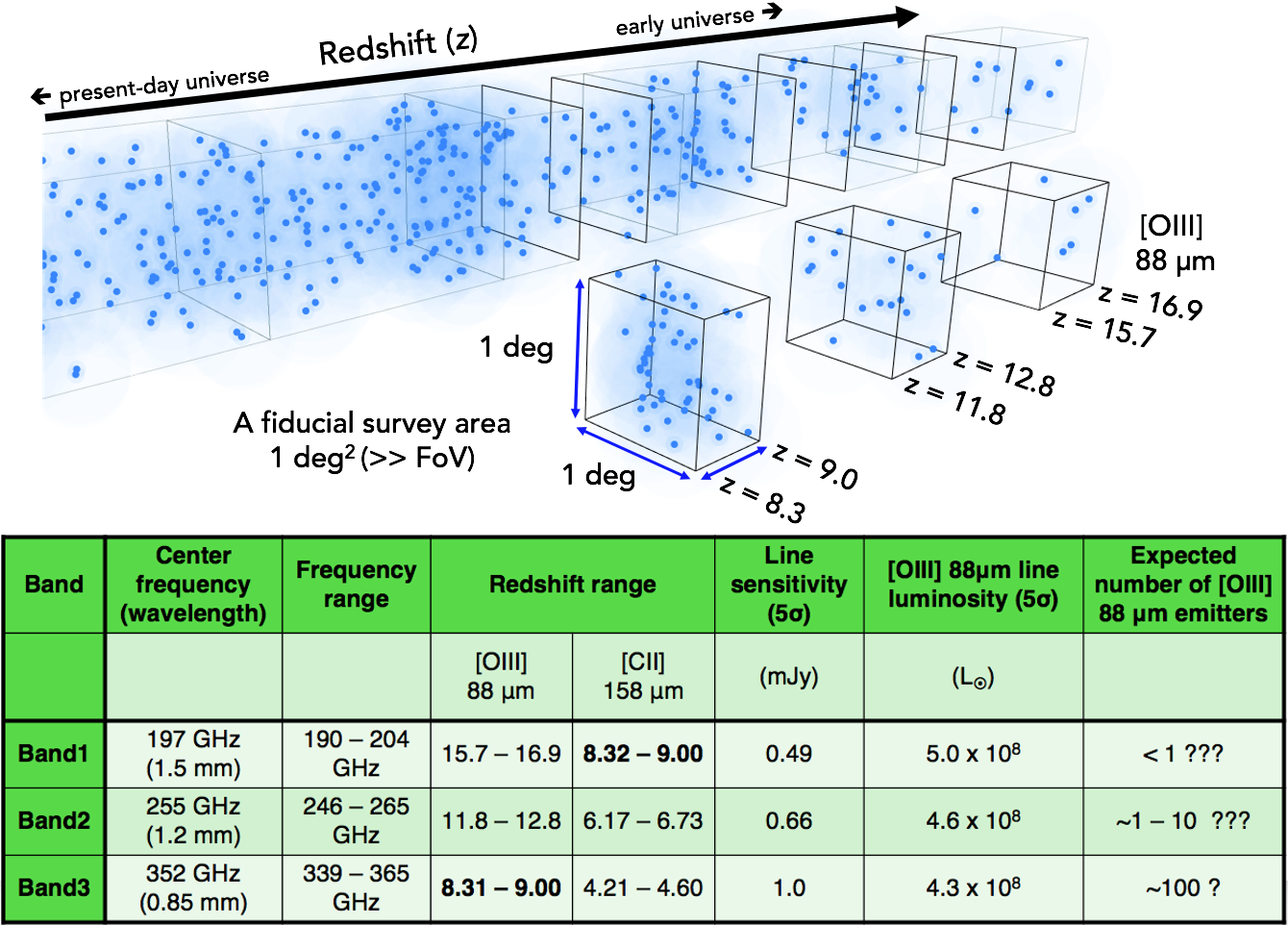}
   \end{tabular}
   \end{center}
   \caption[example] 
   { \label{fig:survey} 
   The proposed 1 deg$^2$ drilling survey using the KATANA 3-band imaging spectrograph, which is designed to blindly uncover \oiii 88 $\mu$m line emitters at 3 specific redshift ranges. The lowest frequency band also corresponds to a \cii 158 $\mu$m redshift range of $z$ = 8.3--9.0, allowing to detect \oiii - \cii dual line emitters at this redshift range. With the observing time of 3,500 hrs, we will reach a 5$\sigma$ line sensitivity of 0.5--1.0 mJy (peak flux) or 0.074--0.15 Jy km s$^{-1}$, which are translated into a \oiii 88 $\mu$m line luminosity of $\sim(4-5)\times10^8$ $L_\odot$. Note that with this line sensitivity we can detect both \oiii and \cii lines of MACS0416\_Y1 at $z=8.31$ \cite{Tamura2019,Bakx2020}.
   }
   \end{figure}

We propose a 3-band (200, 255, and 350 GHz), medium resolution ($R$=2,000) imaging spectrograph with $\sim$1.5 M detectors in total, which is anticipated to become available in the early 2030s. The key technologies of the the KATANA concept (Karatsu et al.~2019) are a dual-polarization sensitive broadband antenna, an on-chip filter-bank spectrometer, and NbTiN-Al hybrid microwave kinetic inductance detectors (MKIDs) to readout the spectral channels, which have been successfully demonstrated by the recent DESHIMA 1.0 on ASTE campaign \cite{Endo2019,Endo2019NA,Takekoshi2020}. DESHIMA 2.0, which will have a much wider frequency coverage with a better optical efficiency, will be deployed in ASTE for full science operation in mid-2021. MOSAIC, a multi-pixel version of DESHIMA, has been proposed for the Large Millimeter Telescope Alfonso Serrano (LMT)\footnote{\url{http://lmtgtm.org}} and discussed during the Guillermo Haro 2018 Workshop.  
Figure \ref{fig:deshima2katana} presents a summary of major specifications of DESHIMA, MOSAIC, and the proposed 3-band imaging spectrograph with the KATANA concept. Here ``KATANA'' is for the existing facilities like ASTE and LMT, whereas KATANA 2.0 is for the future survey telescopes in the early 2030s, i.e., the LST/AtLAST. Figure \ref{fig:survey} summarizes the frequency ranges and corresponding redshifts of \oiii and \cii, along with the conceptual view of a fiducial 1-deg$^2$ drilling survey.  

One of the key requirements for the instrument is a spectral resolution $R$ of 2,000 (a velocity resolution $dv$ of 150 km s$^{-1}$), because the target sources at $z>8$ are expected to be less massive and therefore have a narrow line width of $\sim$100 km s$^{-1}$, as demonstrated by reported \oiii 88 $\mu$m line emitters at $z$ = 7--9 ($dv$ = a few 10--150 km s$^{-1}$)\cite{Inoue2016,Hashimoto2018,Tamura2019}. This is currently the limiting factor of the spatial pixels because we need $>$100 spectral channels to have a redshift width $dz$ of around unity for each band (Figure \ref{fig:survey}). 
The line mapping speed of the proposed configurations was computed based on the achievements with the DESHIMA 1.0 along with the reasonable assumptions of the telescope surface accuracy (45 $\mu$m rms) and a precipitable water vapor (0.5 mm). 
We set a 5$\sigma$ line sensitivity of 1 mJy (peak flux) or 0.15 Jy km s$^{-1}$ at Band-3, which corresponds to a \oiii 88 $\mu$m line luminosity of $4\times10^8$ $L_\odot$ at $z\sim$ 8--9, based on the estimation of the \oiii 88 $\mu$m line luminosity functions from the UV luminosity functions\cite{Oesch2018,Ishigaki2018} and their extrapolation (Inoue et al.). Theoretical simulations\cite{Moriwaki2018} also support the validity of the target line sensitivity. 

We find that a 1-deg$^2$ survey spending 3,500 hrs with the proposed 3-band configuration will uncover a statistically large number of \oiii 88 $\mu$m line emitters at $z$ = 8--9. A fraction of them will be \oiii - \cii dual line emitters because \cii can also be bright at $z\sim8$ galaxies\cite{Bakx2020}. Furthermore, Band-2 is designed to cover the same redshift range in the \nii 122 $\mu$m line. Although the nitrogen line must be significantly weaker than \oiii and \cii, stacking may work to assess the average ISM properties, such as metallicity, radiation field, and gas density\cite{Tadaki2019}. The expected number of sources for the redshift bins of $\sim12$ and 16 is highly uncertain at this stage, but we will be able to put a meaningful observational constraint on the \oiii 88 $\mu$m luminosity functions at $z\sim12$ and 16. A wider survey with this depth will be necessary for better statistics in any case, implying for the necessity for more detectors to provide a higher line mapping speed.

\section{Technical considerations}

\subsection{ISS option}

One of the technical issues is how to realize $R$ = 2,000 with the ISS technology. DESHIMA1.0 has a coplanar waveguide (CPW) signal line coupled to planar filters, but suffers from large radiation losses. Microstrip filters will eliminate such losses, but then we need transparent deposited dielectrics. The development will require a better understanding of the noise behavior of a deposited dielectric, which has two level systems (TLS) noise\cite{Bruno2020TLS}. 

Another challenge is how to implement $>$1,000 spatial pixels which are coupled with a broad-band antenna. Further investigations on the quasi-optics as well as the chip structure will be necessary to have a better design for very large arrays. 

The line mapping speed with the proposed 3-band configuration, which requires $\sim$1.5 M detectors in total, seems to be minimal to achieve the science goal with a reasonable observing time (for $z = 12$ and 16, we need an even wider surveys with this depth).
It implies that the cost of the proposed imaging spectrograph can be comparable to the telescope itself. What is the anticipated cost per channel in 2030s?

\subsection{Heterodyne option}

High spectral resolution ($R \sim 10^7$ or $dv$ $\sim$0.03 km s$^{-1}$) is available, but such high $R$ may not be mandatory for the proposed high-redshift galaxy survey using emission lines; a moderate $R$ (a few 1,000) can work. Nevertheless, the heterodyne receivers are indeed an attractive option as demonstrated by a 275--500 GHz heterodyne receiver with an instantaneous IF bandwidth of 4--21 GHz\cite{Kojima2020}, which exploits the high-current-density superconductor-insulator-superconductor (SIS) junctions and related technologies. It is therefore conceivable to develop a dedicated moderate-resolution digital spectrometer system to be connected to such a novel wide-band heterodyne receiver. 

Another technical challenge is implementation of $\sim10^3$ pixel heterodyne receiver array, where power consumption of a large number of cryogenic low-noise amplifiers and complexity of the structure can be an issue, although possible solutions have been proposed and investigated\cite{Uzawa2018,Shan2020}. 
A heterodyne receiver array with $\sim$1,000 pixels per band has been considered for the AtLAST, and the primary limiting factor is likely to be cost\cite{Groppi2019}.

\section{Summary and outlook}

Successive detection of the \oiii 88 $\mu$m line in $z$ = 7--9 galaxies along with the 
mounting evidence for the earliest star formation at $z\sim$12--15 galaxies motivate us to establish a pathway to find suitable targets for ALMA spectroscopy at these redshift range. We propose and discuss a 3-band imaging spectroscopy survey with the KATANA concept (Karatsu et al. 2019) for LST/AtLAST by exploiting the technologies of the integrated superconducting spectrometer (ISS) and a large-format imaging array. 
A 1-deg$^2$ drilling survey (3,500 hrs) with the proposed 3-band configuration (covering ~200, 255, and 350 GHz band, 1,000--3,000 spatial pixels per band, $R$ = 2,000 spectroscopy, yielding $\sim$1.5 M detectors in total) will capture a statistically large number of \oiii line emitters at $z\sim$ 8--9 (with some \oiii - \cii dual line detection along with \nii 122 $\mu$m constraint by stacking), and a significant chance of uncovering \oiii emitter candidates up to at $z \sim$12 and $\sim$16.
Development of low-loss films, which requires better understanding of noise behavior of dielectrics, is necessary for realizing $R$ = 2,000 with ISS. How to implement $>$1,000 spatial pixels with a broad-band antenna is another challenge.
A large format heterodyne array may also become an option given the rapid progress of SIS receiver technologies, but the cost remains a big issue.

\acknowledgements

This research was supported by 
the Netherlands Organization for Scientific Research NWO (Vidi grant no.~639.042.423, NWO Medium Investment grant no.~614.061.611 DESHIMA), 
the European Research Council ERC (ERC-CoG-2014
- Proposal no.~648135 MOSAIC), 
the Japan Society for the Promotion of Science JSPS (KAKENHI grant nos. JP25247019 and JP17H06130), 
National Astronomical Observatory of Japan NAOJ ALMA Scientific Research (grant no.~2018-09B), 
NAOJ Research Coordination Committee, National Institutes of Natural Sciences (grant no.~1901-0102), and 
the Grant for Joint Research Program of the Institute of Low Temperature Science, Hokkaido University.

\bibliography{report} 
\bibliographystyle{spiebib} 

\end{document}